\documentclass[pdflatex,sn-mathphys-num]{sn-jnl}

\usepackage{graphicx}%
\usepackage{multirow}%
\usepackage{amsmath,amssymb,amsfonts}%
\usepackage{amsthm}%
\usepackage{mathrsfs}%
\usepackage[title]{appendix}%
\usepackage{xcolor}%
\usepackage{textcomp}%
\usepackage{manyfoot}%
\usepackage{booktabs}%
\usepackage{algorithm}%
\usepackage{algorithmicx}%
\usepackage{algpseudocode}%
\usepackage{listings}%

\theoremstyle{thmstyleone}%
\theoremstyle{thmstyletwo}%

\theoremstyle{thmstylethree}%

\begin{document}

\title[Pathology-Guided Virtual Staining Metric for Evaluation and Training]{Pathology-Guided Virtual Staining Metric for Evaluation and Training}


\author[1]{\fnm{Qiankai} \sur{Wang}}\email{qiankai.wang@uwaterloo.ca}

\author[1]{\fnm{James E.D.} \sur{Tweel}}\email{james.tweel@uwaterloo.ca}

\author*[1]{\fnm{Parsin} \sur{Haji Reza}}\email{parsin.hajireza@uwaterloo.ca}
\equalcont{These authors contributed equally to this work.}

\author[2,3,4,5]{\fnm{Anita} \sur{Layton}}\email{anita.layton@uwaterloo.ca}
\equalcont{These authors contributed equally to this work.}

\affil*[1]{\orgdiv{Department of Systems Design Engineering}, \orgname{University of Waterloo}, \orgaddress{\street{200 University Ave W}, \city{Waterloo}, \postcode{N2L 3G1}, \state{Ontario}, \country{Canada}}}

\affil[2]{\orgdiv{Department of Applied
Mathematics}, \orgname{University of Waterloo}, \orgaddress{\street{200 University Ave W}, \city{Waterloo}, \postcode{N2L 3G1}, \state{Ontario}, \country{Canada}}}

\affil[3]{\orgdiv{Cheriton School of Computer
Science}, \orgname{University of Waterloo}, \orgaddress{\street{200 University Ave W}, \city{Waterloo}, \postcode{N2L 3G1}, \state{Ontario}, \country{Canada}}}

\affil[4]{\orgdiv{Department of Biology}, \orgname{University of Waterloo}, \orgaddress{\street{200 University Ave W}, \city{Waterloo}, \postcode{N2L 3G1}, \state{Ontario}, \country{Canada}}}

\affil[5]{\orgdiv{School of Pharmacy}, \orgname{University of Waterloo}, \orgaddress{\street{10 Victoria St S A}, \city{Kitchener}, \postcode{N2G 1C5}, \state{Ontario}, \country{Canada}}}


\abstract{Virtual staining has emerged as a powerful alternative to traditional histopathological staining techniques, enabling rapid, reagent-free image transformations. However, existing evaluation methods predominantly rely on full-reference image quality assessment (FR-IQA) metrics such as structural similarity, which are originally designed for natural images and often fail to capture pathology-relevant features. Expert pathology reviews have also been used, but they are inherently subjective and time-consuming.

In this study, we introduce PaPIS (Pathology-Aware Perceptual Image Similarity), a novel FR-IQA metric specifically tailored for virtual staining evaluation. PaPIS leverages deep learning-based features trained on cell morphology segmentation and incorporates Retinex-inspired feature decomposition to better reflect histological perceptual quality. Comparative experiments demonstrate that PaPIS more accurately aligns with pathology-relevant visual cues and distinguishes subtle cellular structures that traditional and existing perceptual metrics tend to overlook. Furthermore, integrating PaPIS as a guiding loss function in a virtual staining model leads to improved histological fidelity.

This work highlights the critical need for pathology-aware evaluation frameworks to advance the development and clinical readiness of virtual staining technologies.}




\maketitle

\section{Introduction}

In recent years, numerous virtual staining methods have been developed as an alternative to traditional histopathological staining techniques such as hematoxylin and eosin (H\&E), immunohistochemistry (IHC), and Masson’s trichrome \cite{bai2023deep}. Conventional staining methods rely on chemical reagents, which can cause irreversible tissue alterations, limiting subsequent analyses and multi-modal imaging studies. In contrast, virtual staining preserves the tissue in its original state while computationally generating stained representations, enabling repeated analyses and facilitating the integration of different imaging modalities \cite{bai2023deep}.

One of the key advantages of virtual staining is the elimination of chemical reagent consumption. Traditional staining protocols require costly chemical reagents, whereas virtual staining is purely computational, significantly reducing the expense of histopathological workflows. Additionally, virtual staining dramatically accelerates the staining process. While conventional staining can take tens of minutes to several hours, virtual staining can be completed within seconds or minutes, greatly improving the efficiency of pathology pipelines\cite{latonen2024virtual}.

Despite these advantages, virtual staining currently lacks evaluation methods specifically tailored to the medical domain. Most assessments rely on full-reference image quality assessment (FR-IQA) metrics, which compare a processed image to a reference standard based on structural and perceptual similarities. While FR-IQA metrics are well-established in general image processing and have been widely applied in natural image quality assessment, their effectiveness in microscopic pathology imaging remains limited \cite{breger2024studya}. A fundamental limitation of traditional FR-IQA approaches is their focus on textural fidelity rather than histopathological relevance. These metrics are primarily designed for natural images captured by cameras, prioritizing structural and perceptual consistency without considering cellular morphology and tissue architecture, which are crucial for medical applications \cite{breger2024studya, pambrun2015limitations}. Consequently, conventional FR-IQA metrics often fail to accurately assess the diagnostic quality of virtual staining images.

With advances in deep learning and feature engineering, perceptual evaluation metrics have increasingly enabled domain-specific comparisons \cite{zhang2018unreasonable, ding2020image} . However, existing perceptual similarity metrics remain heavily biased toward natural image characteristics, making them inadequate for assessing medical imaging data. As a result, many virtual staining images cannot be effectively evaluated using current methodologies, highlighting the need for a domain-specific, histopathology-aware evaluation framework \cite{breger2024studyb}.

To address this gap, this study proposes a histopathology-guided perceptual full-reference image assessment metric specifically designed for virtual staining evaluation. Additionally, an optimization framework for virtual staining models is developed based on the proposed evaluation metric. By integrating domain-specific histopathological knowledge into image quality assessment, this approach aims to provide a more clinically relevant evaluation of virtual staining results, ultimately improving their reliability and applicability in medical practice.

\section{Literature Review}

Virtual staining has emerged as a powerful approach for synthesizing H\&E-equivalent images from label-free modalities, offering a reagent-free alternative to traditional histopathological staining. This label-free to stain transformation aims to replace physical dyes by generating stained-like outputs directly from inputs such as autofluorescence or photon-based signals. Ecclestone et al.~\cite{ecclestone2024photon} introduced the Photon Absorption Remote Sensing (PARS) system, which captures spectral and temporal photon absorption signatures to emulate H\&E staining with enhanced cellular detail. Building on this direction, Rivenson et al.~\cite{rivenson2018deep} and Wang et al.~\cite{wang2025value} demonstrated the efficacy of deep learning in mapping autofluorescence images to realistic H\&E counterparts, broadening the applicability of virtual staining in clinical workflows.

The advent of generative deep learning models has further propelled the quality and fidelity of virtual staining. Extensions of the PARS framework by Tweel et al.~\cite{tweel2023virtual} and Boktor et al.~\cite{boktor2024multi} integrated generative networks to process expanded spectral inputs and time-domain signals, improving visual realism and structural preservation. In parallel, diffusion-based models, such as those proposed by Saharia et al.~\cite{saharia2022palette}, have shown strong potential for high-fidelity image-to-image translation in biomedical contexts. These advances build on foundational work in generative translation, notably Pix2Pix~\cite{isola2017image} and CycleGAN~\cite{zhu2017unpaired}, which have been adapted to address the specific challenges of virtual staining tasks.

FR-IQA methods compare a processed image to a reference image to quantify differences in structure, perception, and statistical features. Wang et al. \cite{wang2004image} introduced structural similarity index (SSIM), which evaluate pixel-wise differences and structural information to measure image fidelity. To further enhance structural evaluations, Wang et al. \cite{wang2003multiscale} developed multi-scale structural similarity (MS-SSIM), incorporating multi-resolution analysis to improve robustness. Beyond hand-crafted approaches, perceptual-based FR-IQA methods have been introduced to align more closely with human perception. Zhang et al. \cite{zhang2018unreasonable} proposed the Learned Perceptual Image Patch Similarity (LPIPS) metric, which utilizes deep neural networks to model perceptual judgments based on feature representations from trained vision models. Ding et al. \cite{ding2020image} extended perceptual similarity analysis through Deep Image Structure and Texture Similarity (DISTS), integrating structural and texture information to improve quality assessment. Tian et al. \cite{tian2023towards} explored the impact of multiple reference images in quality assessment, while Xian et al. \cite{xian2023perceptual} proposed structure-aware methods utilizing high-order statistical moments to capture intricate quality attributes.

Recent research has explored the application of FR-IQA in medical imaging. Breger et al. \cite{breger2024studya} demonstrated that standard FR-IQA metrics, originally designed for natural images, may not be directly applicable to medical imaging tasks, including MRI, CT, OCT, and digital pathology. Ohashi et al. \cite{ohashi2023applicability} evaluated and adapted FR-IQA methods to better align with medical image quality requirements. Varga et al. \cite{varga2022full} proposed optimized metric combinations to enhance prediction accuracy in medical imaging contexts. Additionally, Sujana et al. \cite{sujana2024full} investigated FR-IQA for structural MRI preprocessing, while Rodrigues et al. \cite{rodrigues2024objective} examined perceptual quality assessment of medical images and videos.

\section{Methods}

\begin{figure}[htbp]
    \centering
    \includegraphics[width=\columnwidth]{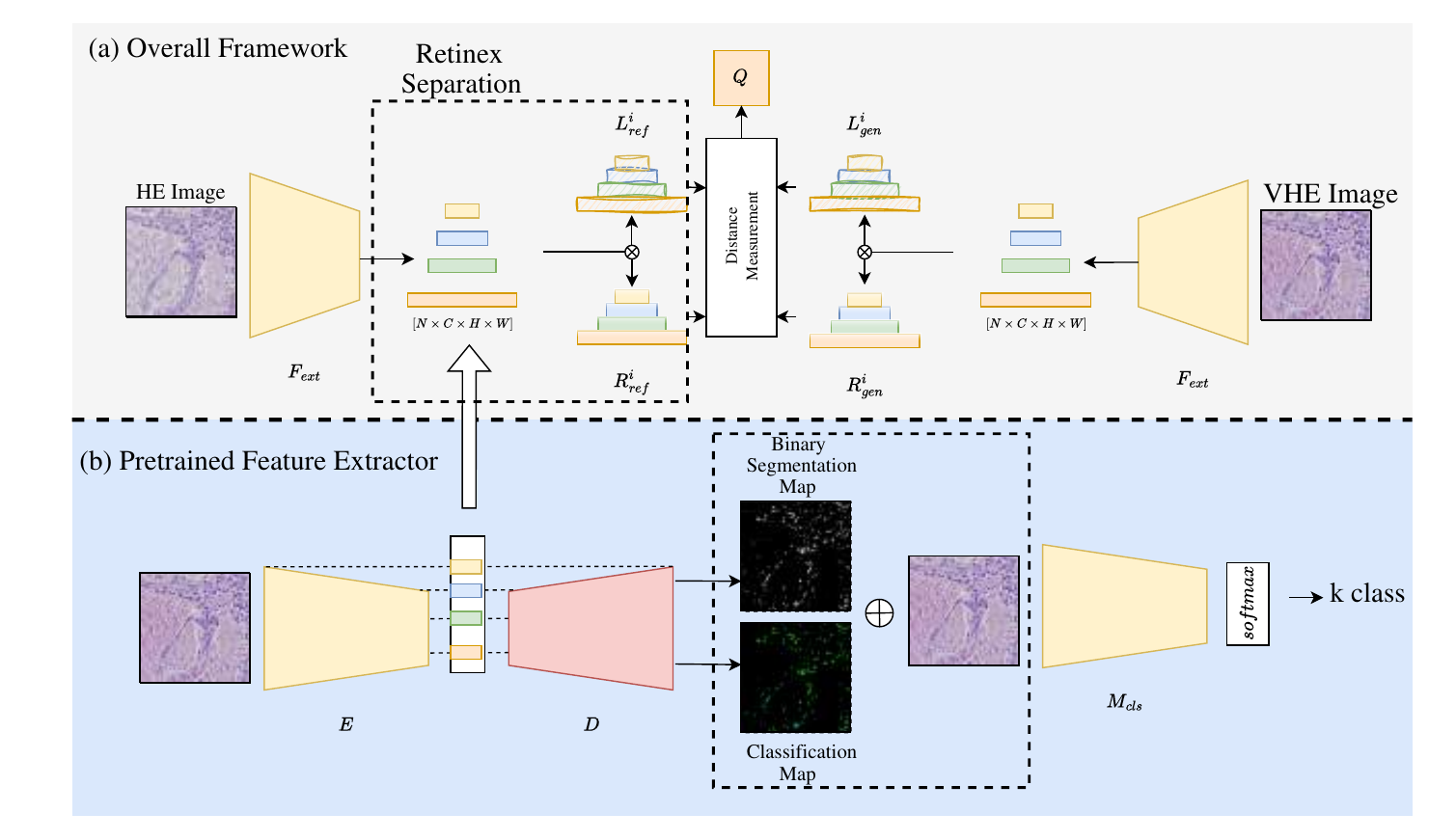}
    \caption{(a) The figure illustrates the overall framework of the PaPIS. Image patches are fed into a pretrained feature extractor to obtain multi-channel feature representations. These features are decomposed into a set of estimated illumination maps \( L^{i}_a \) and reflectance maps \( R^{i}_a \), where the superscript i denotes the feature layer and the subscript \( a \) indicates the image type (Reference or Generated images). The collection of decomposed components is subsequently used to compute perceptual similarity distances between image pairs. (b) This panel presents the internal architecture of the pretrained feature extractor. The feature extractor functions as the encoder component of a cell morphology segmentation model. The subsequent decoder generates both a binary segmentation map \( M_b \), which preserves the spatial dimensions of the input, and a multi-layer classification map \( M_c \). The fusion of \( M_b \), \( M_c \), and the original image patch is then fed into a classification model to produce the final classification label.}
    \label{fig:framework}
\end{figure}

Figure \ref{fig:framework}a illustrates the overall framework of the proposed perceptual similarity metric, PaPIS. This similarity metric assigns perceptual weights that reflect histological performance. To extract feature representations, image patches are first transformed into multi-channel embeddings using a pre-trained cell morphology segmentation model based on the work of Ignatov et al. \cite{ignatov2024histopathological}. Their model, which achieves state-of-the-art performance in nuclei segmentation and classification, adopts a dual-layer encoder-decoder architecture with an EfficientNet-B7-based encoder, as shown in Figure \ref{fig:framework}b.

Subsequently, intrinsic properties are extracted from the image features to obtain the reflection map $R$ and the estimated illumination map $L$, using the Retinex algorithm\cite{jobson1997multiscale}. The histology perceptual distance between these maps is then computed, inspired by prior works such as SSIM \cite{wang2004image} and DISTS \cite{ding2020image}. By calculating the distance between features derived from cell morphology, PaPIS offers a histology-aware quantification of image differences. This contrasts with traditional texture-based metrics such as SSIM and PSNR, which are primarily designed to align with human visual perception rather than pathological relevance.

To validate the practical utility of the proposed metric in pathological image analysis, we further apply it to evaluate the performance of a deep learning model that translates label-free histological images into H\&E-stained representations.

\subsection{Cell Morphology Level Feature Representation}

Current state-of-the-art Full-Reference perceptual image similarity metrics, such as LPIPS\cite{zhang2018unreasonable} and DISTS\cite{ding2020image}, rely on feature extractors pre-trained on ImageNet to evaluate differences between generated and reference images. However, in the histology domain, pathologists focus on distinct perceptual attributes that cannot be adequately captured by models trained on natural image datasets.

To achieve superior performance in histological assessments, we utilize a feature extractor built upon the implementation of a state-of-the-art cell morphology segmentation task \cite{ignatov2024histopathological}, which has demonstrated exceptional accuracy in nuclei segmentation and classification. In work by ignatov et al. \cite{ignatov2024histopathological}, cell morphology segmentation is achieved using a dual-layer encoder-decoder architecture with an EfficientNet-B7-based encoder, the structure of which is shown in Figure \ref{fig:framework}b. The computed latent features utilized in our work, formally, can be written as:

\begin{equation}
f_{eff}(x) = \{\tilde{x}^{(i)}_{j}; i=0, \dots, m; j=1, \dots, n_{i}\}
\end{equation}

where \( x \in \mathbb{R}^{H \times W \times C} \) denotes the input image patch, \( m = 4 \) represents the number of selected convolutional blocks within the EfficientNet backbone, and \( n_i \) is the number of output channels (feature maps) at block \( i \). The term \( \tilde{x}^{(i)}_j \) refers to the \( j \)-th feature map from block \( i \), normalized via channel-wise min-max normalization to ensure scale consistency across layers. 

This formulation explicitly defines the multi-layer, multi-channel output of the encoder, which serves as the foundation for subsequent perceptual analysis. The architecture of the feature extractor is illustrated in Figure~\ref{fig:framework}b.

To compare the perceptual focus of these distinct feature extractors, and obtain a unified representation from the multi-layer feature maps, we compute a reconstructed image \( I_{\text{Rec}} \) as follows:

\begin{equation}
I_{\text{Rec}} = \frac{1}{m} \sum_{i=0}^{m} f_{in} \left( \frac{1}{n} \sum_{j=0}^{n} \tilde{x}^{(i)}_{j} \right)
\end{equation}

Here, \( \tilde{x}^{(i)}_{j} \) denotes the \( j \)-th normalized feature map at layer \( i \), as previously defined. The inner summation computes the average feature response across all channels \( n \) within each layer, while the function \( f_{in}(\cdot) \) integrates these responses. Specifically, \( f_{in} \) denotes a linear interpolation function that maps the current image feature of size \( \mathbb{R}^{w \times h} \) to the original spatial resolution \( \mathbb{R}^{w_0 \times h_0} \), where \( w_0 \) and \( h_0 \) correspond to the width and height of the input image. The outer average aggregates information across all \( m \) selected layers, resulting in the final reconstructed representation \( I_{\text{Rec}} \), which captures both local and hierarchical morphological features from the input image.

\begin{figure}[htbp]
    \centering
    \includegraphics[width=\columnwidth]{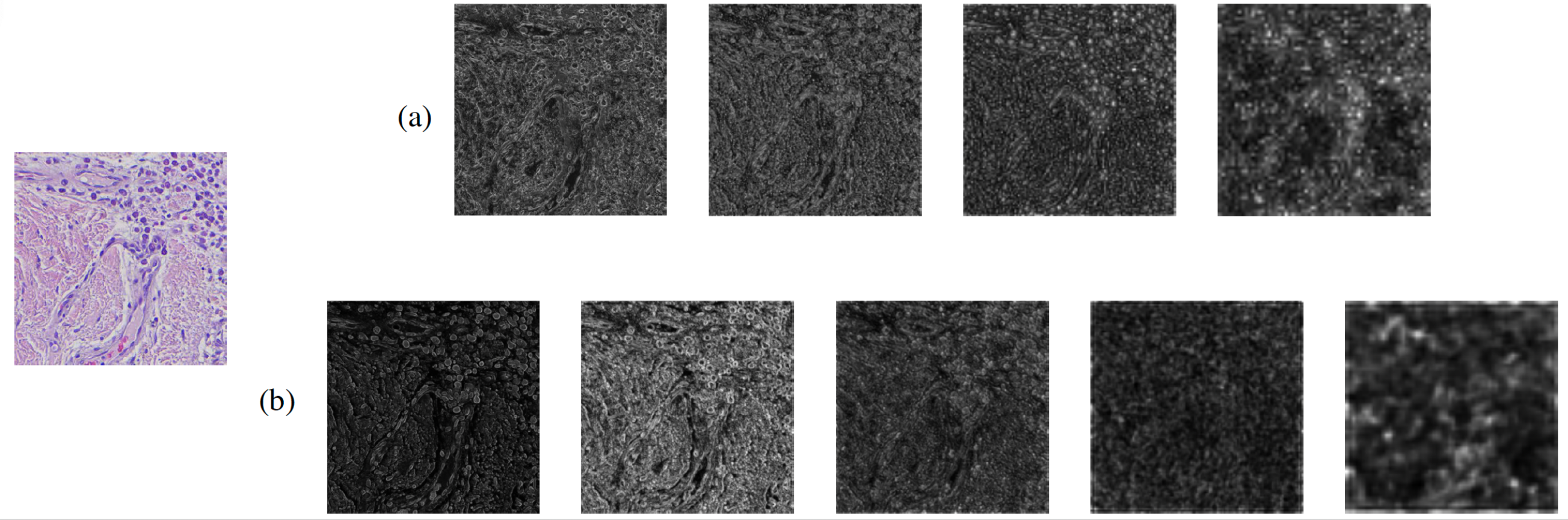}
    \caption{(a) The first set of feature maps was visualized from a VGG16 model pretrained on the ImageNet dataset. These feature maps correspond to channels 3, 8, 15, 22, and 29, as used in the works of LPIPS \cite{zhang2018unreasonable} and DISTS \cite{ding2020image}. (b) The second set of feature maps was extracted from the encoder output of a pretrained segmentation model, with dimensions of $R^{192}$, $R^{288}$, $R^{480}$, and $R^{1344}$, corresponding to channels at different encoder stages.}
    \label{fig:features}
\end{figure}

Figure~\ref{fig:features} illustrates the visualization of extracted features, comparing the feature maps obtained from models pre-trained on the cell morphology histological task (\( F_{\text{histo}} \)) with those from models pre-trained on natural image datasets (\( F_{\text{natural}} \)). Specifically, Figure~\ref{fig:features}b shows that the deeper layers of \( F_{\text{histo}} \) focus more strongly on nuclear regions, exhibiting activation patterns that correspond to the locations of cell nuclei. In contrast, Figure~\ref{fig:features}a demonstrates that \( F_{\text{natural}} \) primarily captures low-level texture patterns such as edges and gradients, which are characteristic of natural images and reflect the visual features typically emphasized by models trained to match human perception.

\subsection{Retinex Properties on Features}

In order to enhance the contrast of histological features, we apply the Multi-Scale Retinex (MSR) algorithm~\cite{jobson1997multiscale} to decompose the extracted features into two distinct components: the estimated illumination map and the reflectance map, which correspond to the low-frequency and high-frequency components within the extracted feature space, respectively. The MSR algorithm models the decomposition of an input feature map based on the following relationship:

\begin{equation}
    I(x, y) = R(x, y) \cdot L(x, y)
\end{equation}

where \( I(x, y) \) denotes the observed feature intensity at pixel location \( (x, y) \), \( R(x, y) \) is the reflectance component capturing intrinsic feature structures (e.g., edges and nuclei boundaries), and \( L(x, y) \) represents the illumination component modeling smooth variations in local intensity due to shading or uneven activation.

To obtain a multi-scale estimation, we compute the illumination component at each scale \( i \) as:

\begin{equation}
    L_i(x, y) = F_i(x, y) * G_{\sigma_i}(x, y)
\end{equation}

where \( F_i(x, y) \) denotes the extracted feature map at scale \( i \), \( G_{\sigma_i}(x, y) \) is a 2D Gaussian filter with standard deviation \( \sigma_i \), and \( * \) denotes the 2D convolution operation. The choice of \( \sigma_i \) controls the spatial frequency captured at each scale.

The reflectance component at each scale is then derived by logarithmic subtraction:

\begin{equation}
    R_i(x, y) = \log F_i(x, y) - \log L_i(x, y)
\end{equation}

where the logarithmic transformation serves to suppress multiplicative illumination effects and enhance the structural contrast in the feature map. The resulting \( R_i(x, y) \) highlights fine-grained morphological features relevant to histological analysis.

\subsection{PaPIS: Pathological Perceptual Distance}

PaPIS integrates both low-frequency and high-frequency perceptual distances to evaluate histopathological image similarity. The decomposition of feature maps into illumination and reflectance components is based on the Retinex theory~\cite{jobson1997multiscale}, which models an image as the product of its intrinsic reflectance and smooth illumination.

Inspired by SSIM~\cite{wang2004image} and DISTS~\cite{ding2020image}, we define the high-frequency distance by comparing reflectance features extracted from Retinex decomposition. The high-frequency distance, denoted as \( D_{\text{high}}(x, y, \alpha, \beta) \), is formulated as:

\
\begin{equation} \label{eq:highd}
\begin{split}
D_{\text{high}}(x, y, \alpha, \beta) =\; & \alpha_{ij} \frac{2\mu^{(i)}_{R \tilde{x}_{j}}\mu^{(i)}_{R \tilde{y}_{j}}+c_{1}}{(\mu^{(i)}_{R \tilde{x}_{j}})^2 + (\mu^{(i)}_{R \tilde{y}_{j}})^2+c_{1}} \\
& + \beta_{ij} \frac{2\sigma^{(i)}_{R \tilde{x}_{j}}\sigma^{(i)}_{R \tilde{y}_{j}}+c_{2}}{(\sigma^{(i)}_{R \tilde{x}_{j}})^2 + (\sigma^{(i)}_{R \tilde{y}_{j}})^2+c_{2}}
\end{split}
\end{equation}

In this formulation, the first term captures the similarity of mean values (\( \mu \)) of the reflected features, representing brightness consistency across spatial locations and feature channels. The second term evaluates the similarity of standard deviations (\( \sigma \)), which reflects structural consistency in the high-frequency domain. The reflection statistics \( \mu^{(i)}_{R \tilde{x}_{j}} \), \( \mu^{(i)}_{R \tilde{y}_{j}} \), \( \sigma^{(i)}_{R \tilde{x}_{j}} \), and \( \sigma^{(i)}_{R \tilde{y}_{j}} \) denote the mean and standard deviation of reflectance features \( \tilde{x} \) and \( \tilde{y} \) at layer \( i \) and channel \( j \). Constants \( c_1 \) and \( c_2 \) are included to ensure numerical stability. The weights \( \alpha_{ij} \) and \( \beta_{ij} \) are randomized such that \( \sum_{i,j} \alpha_{ij} + \beta_{ij} = 1 \), ensuring a balanced contribution between brightness and structure.

The low-frequency distance, \( D_{\text{low}} \), is computed as the mean squared error (MSE) between illumination maps derived from each feature channel via the Retinex algorithm \cite{jobson1997multiscale}.

Finally, the complete PaPIS metric is defined as:
\begin{equation}
\begin{split}
    PaPIS(x, y, \lambda, \alpha, \beta) =\; & \lambda \sum_{i=0}^{m} \sum_{j=0}^{n_i} 
    \text{MSE}(L(\tilde{x}^{(i)}_{j}), L(\tilde{y}^{(i)}_{j})) \\
    & + D_{\text{high}}(x, y, \alpha, \beta)
\end{split}
\end{equation}

Here, \( \lambda \) is a hyperparameter that balances the contributions of the low-frequency and high-frequency components in the final distance score.

\subsection{PaPIS Metric-Guided Loss for Enhancing Virtual Staining Models} 

\begin{figure}[htbp]
    \centering
    \includegraphics[width=\columnwidth]{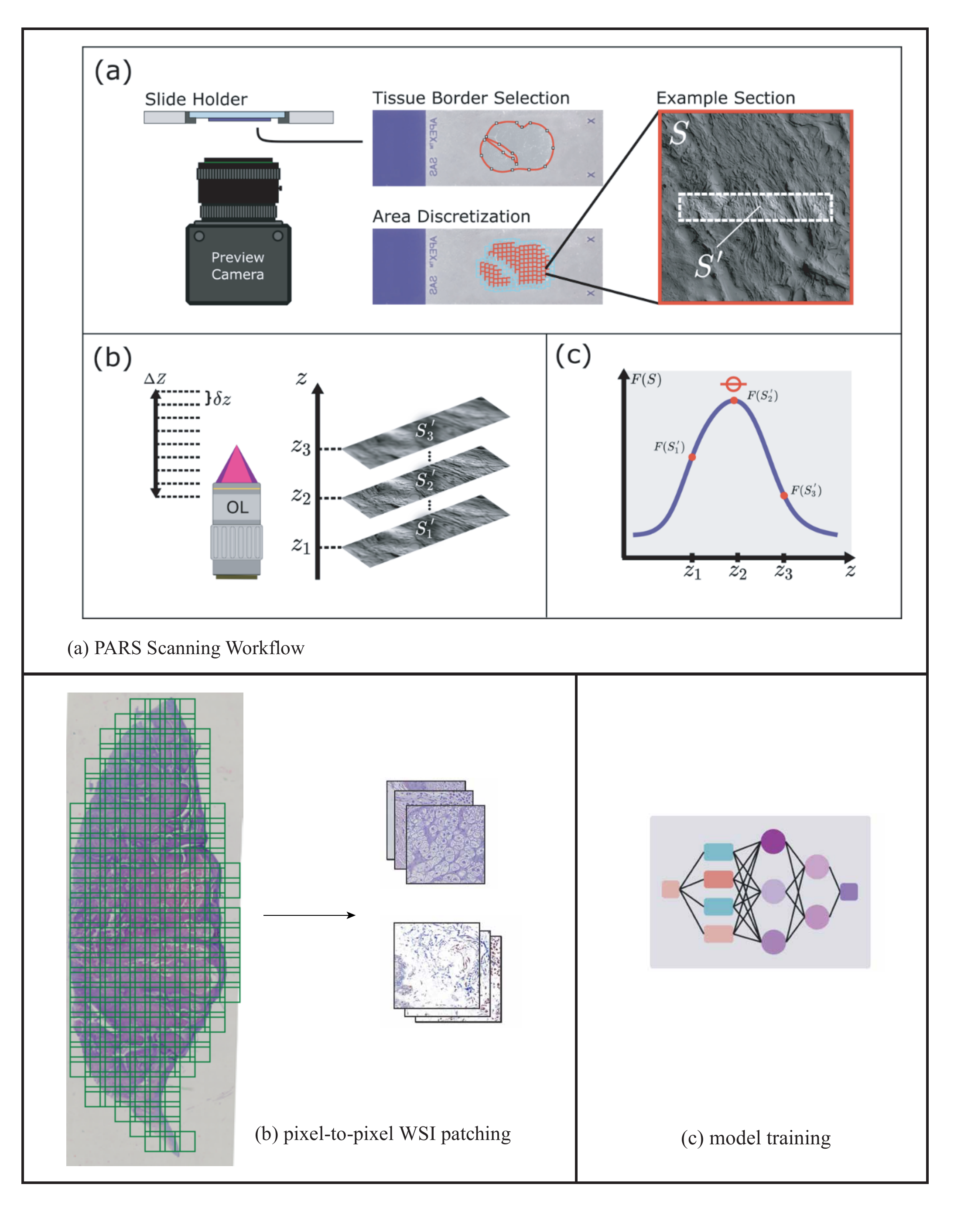}
    \caption{Pipeline for the PaPIS-guided virtual staining. Part (a) is adapted from \cite{tweel2024automated}. The complete pipeline consists of slide scanning, patch extraction, and model training, forming an end-to-end framework for virtual staining.}
    \label{fig:papis-pipeline}
\end{figure}

Since PaPIS is utilized as a quality evaluation metric in generative tasks, incorporating it as an optimization objective during training could further improve the performance of virtual staining. Building on the unpaired training paradigm of the CycleGAN model, we propose a PaPIS-guided framework for virtual staining, designed to transform images from the PARS modality into H\&E-stained representations. The complete pipeline is illustrated in Figure \ref{fig:papis-pipeline}.

\subsubsection{PARS images Acquisition} \label{image-acquisition}

The pixel-by-pixel registration of scanned whole slide images (WSIs) and the enhancement process are based on the work by Tweel et al. \cite{tweel2024automated}. Training patches are extracted using an automated script specifically designed for large-scale whole slide images. This script is capable of extracting the required patches within 15 seconds, significantly improving efficiency. The training dataset is constructed in paired format rather than in a random order, following the methodology outlined in the work of Tweel et al. \cite{tweel2023virtual}.

\subsubsection{PaPIS-guided Virtual Staining CycleGAN}

Building upon the standard CycleGAN framework~\cite{zhu2017unpaired}, we introduce an auxiliary perceptual loss term based on the proposed PaPIS metric to encourage the generated images to better align with histological properties of the target domain. This PaPIS-based loss is defined as:

\begin{equation}
    \mathcal{L}_{\text{papis}}(x, G(x)) = 1 - \text{PaPIS}(x, G(x), \lambda, \alpha, \beta)
\end{equation}

where \( x \) is the input unstained image and \( G(x) \) is the corresponding virtual H\&E image generated by the generator \( G \). The PaPIS score quantifies the pathological perceptual similarity between the input and generated image; thus, minimizing \( \mathcal{L}_{\text{papis}} \) encourages perceptual closeness in both low- and high-frequency histological features.

The overall training loss for the modified CycleGAN is formulated as:

\begin{equation}
    \mathcal{L}_{\text{total}} = \lambda_{1} \mathcal{L}_{\text{cycle}} + \lambda_{2} \mathcal{L}_{\text{papis}} + \lambda_{3} \mathcal{L}_{\text{g}} + \lambda_{4} \mathcal{L}_{\text{d}}
\end{equation}

\begin{figure}[htbp]
    \centering
    \includegraphics[width=\columnwidth]{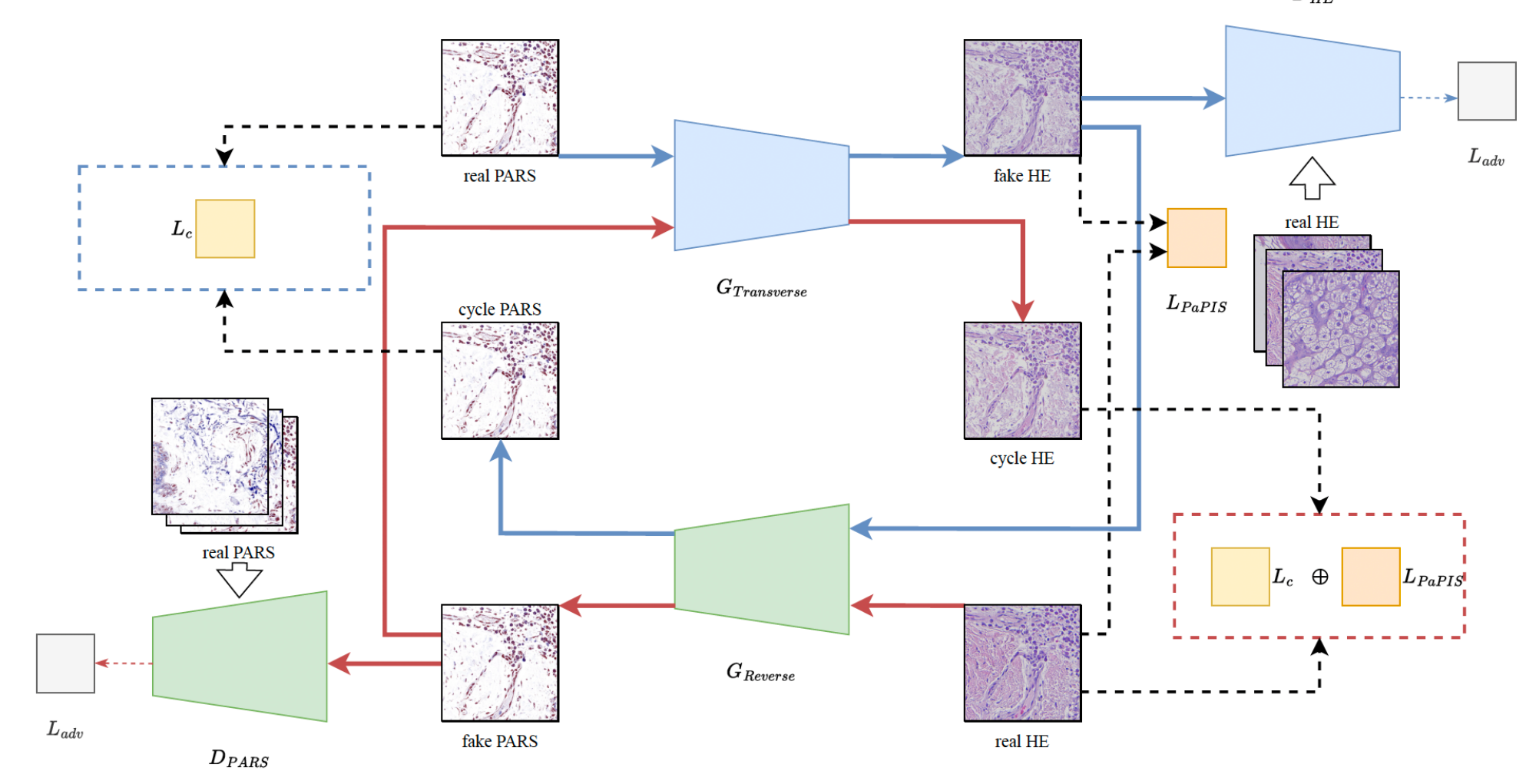}
    \caption{The architecture of the PaPIS-guided virtual staining model. Built upon the CycleGAN\cite{zhu2017unpaired} framework, the model incorporates additional loss terms to measure the distance between the generated (fake) H\&E images and the real H\&E images (ground truth), as well as the distance between cycle-consistent H\&E images and the real H\&E images.}
    \label{fig:cycleGAN}
\end{figure}

Here, \( \mathcal{L}_{\text{cycle}} \) denotes the cycle consistency loss, \( \mathcal{L}_{\text{g}} \) is the generator adversarial loss, and \( \mathcal{L}_{\text{d}} \) is the discriminator loss. The weights \( \lambda_1, \lambda_2, \lambda_3, \lambda_4 \) are hyperparameters that control the contribution of each loss component. Figure~\ref{fig:cycleGAN} illustrates the modified CycleGAN architecture with PaPIS-guided perceptual supervision.

\section{Experiments}

\subsection{Dataset Acquisition}

\subsubsection{Modalities of Images Acquisition}

All scanning and post-processing procedures are detailed in Section~\ref{image-acquisition}. The dataset consists of five pairs of whole slide images (WSIs) with dimensions of \(18,202 \times 48,800\), \(9,200 \times 36,200\), \(59,600 \times 61,418\), \(17,664 \times 17,876\) and \(35,344 \times 37,957\), respectively. Four of the scanned WSIs are divided into \(1024 \times 1024\) patches, resulting in a total of 10,086 patches. These patches are then randomly flipped and resized to \(256 \times 256\) for the training process. The remaining WSI pair is designated for the comparative experiment. A filtering process is first applied to extract histological regions, followed by randomly cropping into \(206\) synchronized \(1024 \times 1024\) patches for subsequent analysis.

\subsubsection{Virtual Stained Image Generate} \label{PaPIS-guided}

The dataset utilized in this study was generated using a virtual staining model based on the CycleGAN architecture \cite{zhu2017unpaired}. The loss function parameters were set as follows: $\lambda_{1} = 2.0$, $\lambda_{2} = 1.0$, $\lambda_{3} = 1.0$, and $\lambda_{4} = 1.0$.
Training was conducted with a batch size of 1. The Adam optimizer was used for both the generator and discriminator networks, with a learning rate of 0.001 and $\beta$ parameters of $(0.5, 0.999)$. An accumulative gradient strategy with an accumulation count of 2 was applied.
A linear learning rate decay strategy was employed using the \texttt{LinearLrInterval} scheduler, with updates every 1000 iterations. The learning rate decayed linearly from 0.001 to 0 between the 10,000\textsuperscript{th} and 50,000\textsuperscript{th} iterations.

\subsection{Comparison with Previous Metrics}

For the comparison of image pairs, we calculate the following metrics: PSNR, SSIM \cite{wang2004image}, MS-SSIM \cite{wang2003multiscale}, LPIPS \cite{zhang2018unreasonable}, and DISTS \cite{ding2020image}. Based on the relationship between these metrics and PaPIS scores, we define four distinct categories:
\begin{enumerate}
    \item \textbf{AH (Aligned High)}: Points exhibiting both high PaPIS scores and high values across the compared metrics, indicating strong agreement between PaPIS and traditional measures.
    \item \textbf{AL (Aligned Low)}: Points with both low PaPIS scores and low metric values, reflecting consistent low similarity across both evaluation methods.
    \item \textbf{PD (PaPIS-Dominant)}: Points where PaPIS scores are high despite low traditional metric values, suggesting cases where PaPIS captures relevant pathological features overlooked by conventional metrics.
    \item \textbf{TD (Traditionally-Dominant)}: Points with low PaPIS scores but high traditional metric values, highlighting instances where traditional metrics indicate high similarity, yet PaPIS identifies deficiencies in pathology-relevant features.
\end{enumerate}

We categorize the comparison metrics into two groups: traditional handcrafted metrics and perceptual feature-based perceptual metrics.

\subsubsection{PaPIS and Handcrafted Metrics}

\begin{figure}[htbp]
    \centering
    \includegraphics[width=\columnwidth]{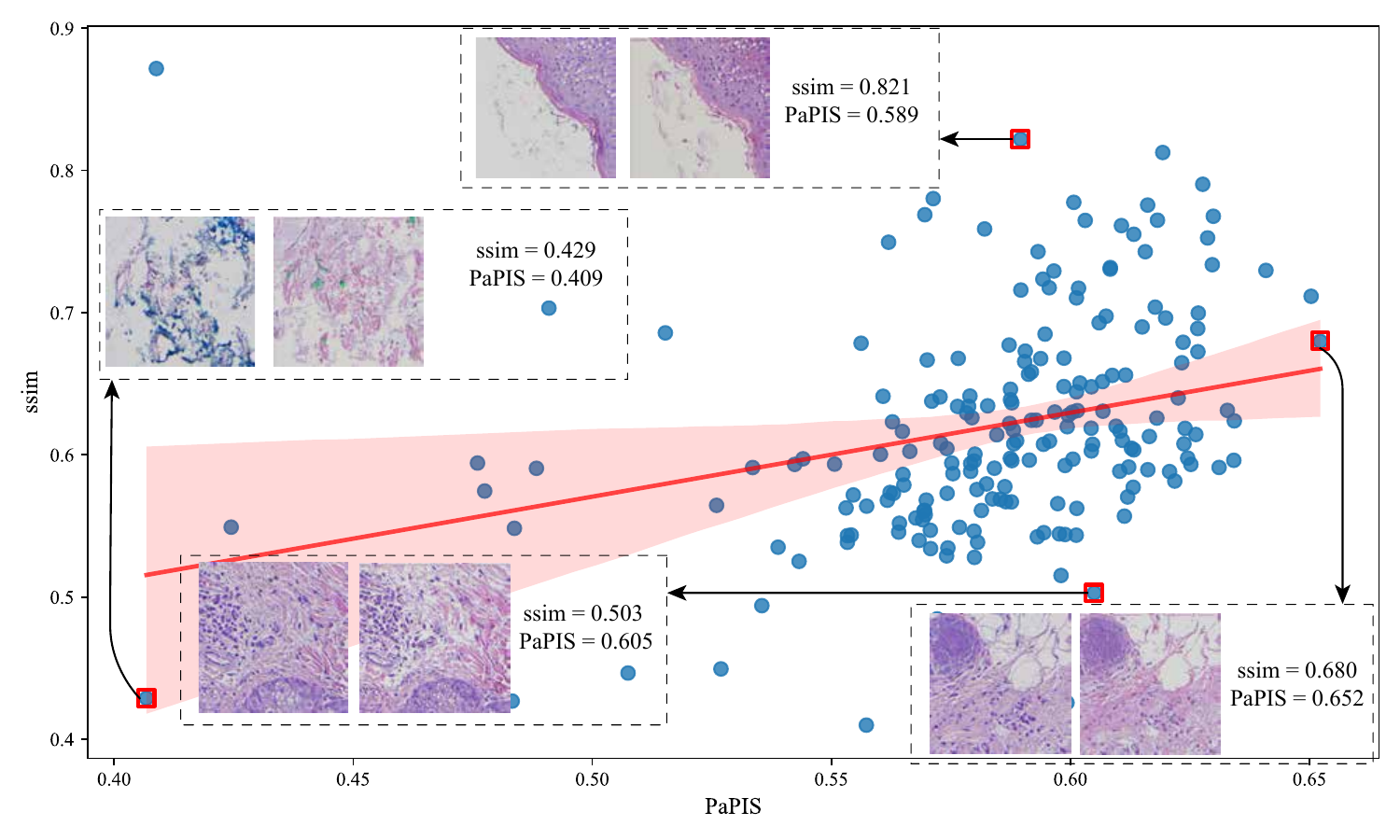}
    \caption{Scatter plot comparing SSIM and PaPIS across different image pairs. Selected data points are labeled as AH (Aligned High), AL (Aligned Low), PD (PaPIS-Dominant), and TD (Traditionally-Dominant), representing specific cases for analysis.}
    \label{fig:scatter_plot_ssim}
\end{figure}

\begin{figure}[htbp]
    \centering
    \includegraphics[width=\columnwidth]{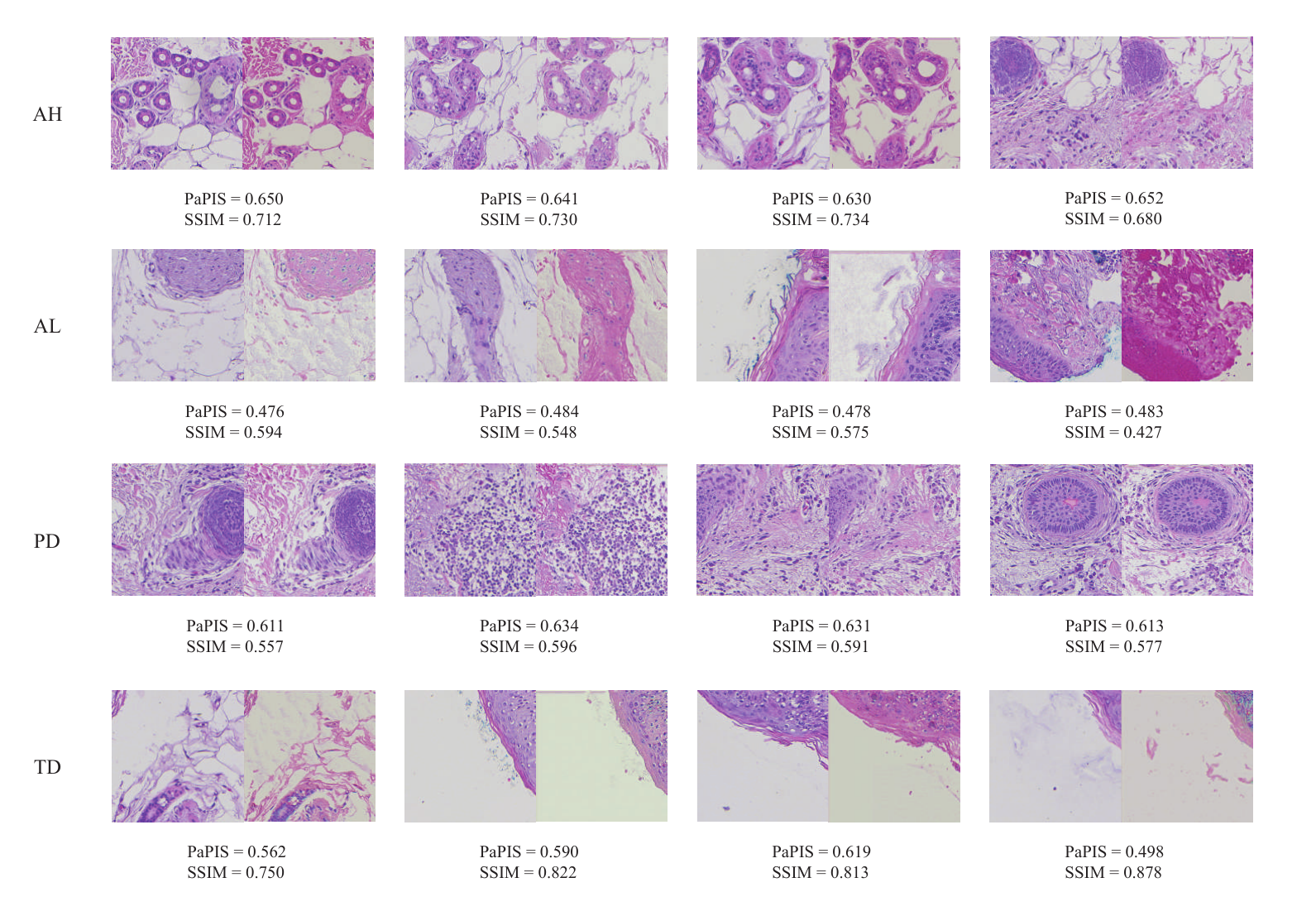}
    \caption{Case-wise comparison of PaPIS and SSIM values across different histological image samples.}
    \label{fig:caselist-papis-ssim}
\end{figure}

In this section, we compare our proposed metric with traditional handcrafted metrics, specifically SSIM \cite{wang2004image}. A case-by-case scatter plot is presented in Figure \ref{fig:scatter_plot_ssim}, while additional image-to-image examples are provided in Figure \ref{fig:caselist-papis-ssim} for further comparison.

Among the AH and AL points, both selected data points exhibit strong alignment between PaPIS and traditional metrics in assessing image quality. In contrast, the PD and TD points highlight the differing sensitivities of PaPIS and traditional approaches. PD points correspond to cases where nuclear structures and spatial alignment are well-preserved, yet fine textures and subtle color variations result in lower scores from traditional metrics, revealing their limitations in capturing histological relevance. Conversely, TD points receive high scores from traditional metrics due to uniform textures or consistent coloration but may lack proper histological structures or exhibit misalignment in cellular organization, leading to lower PaPIS scores. This contrast underscores the ability of PaPIS to better capture pathologically significant features that traditional methods may overlook.

\subsubsection{PaPIS and Other Feature-Based Perceptual Metrics}

\begin{figure}[htbp]
    \centering
    \includegraphics[width=\columnwidth]{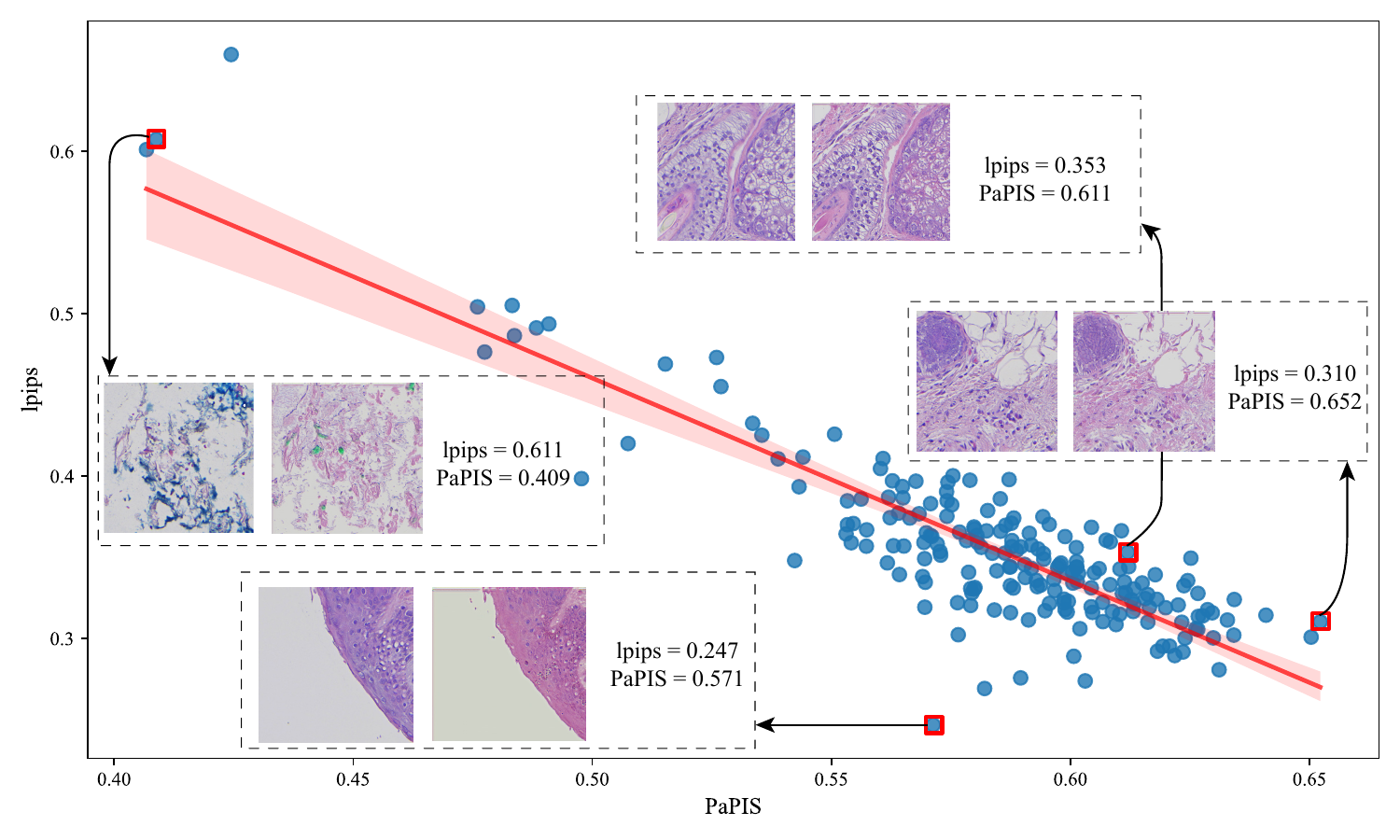}
    \caption{Scatter plot comparing LPIPS and PaPIS across different image pairs. Each data point represents an image pair evaluated by both metrics, illustrating variations in their assessments. This comparison highlights differences in perceptual and pathology-aware similarity measurements between the two approaches.}
    \label{fig:scatter_plot_lpips}
\end{figure}

\begin{figure}[htbp]
    \centering
    \includegraphics[width=\columnwidth]{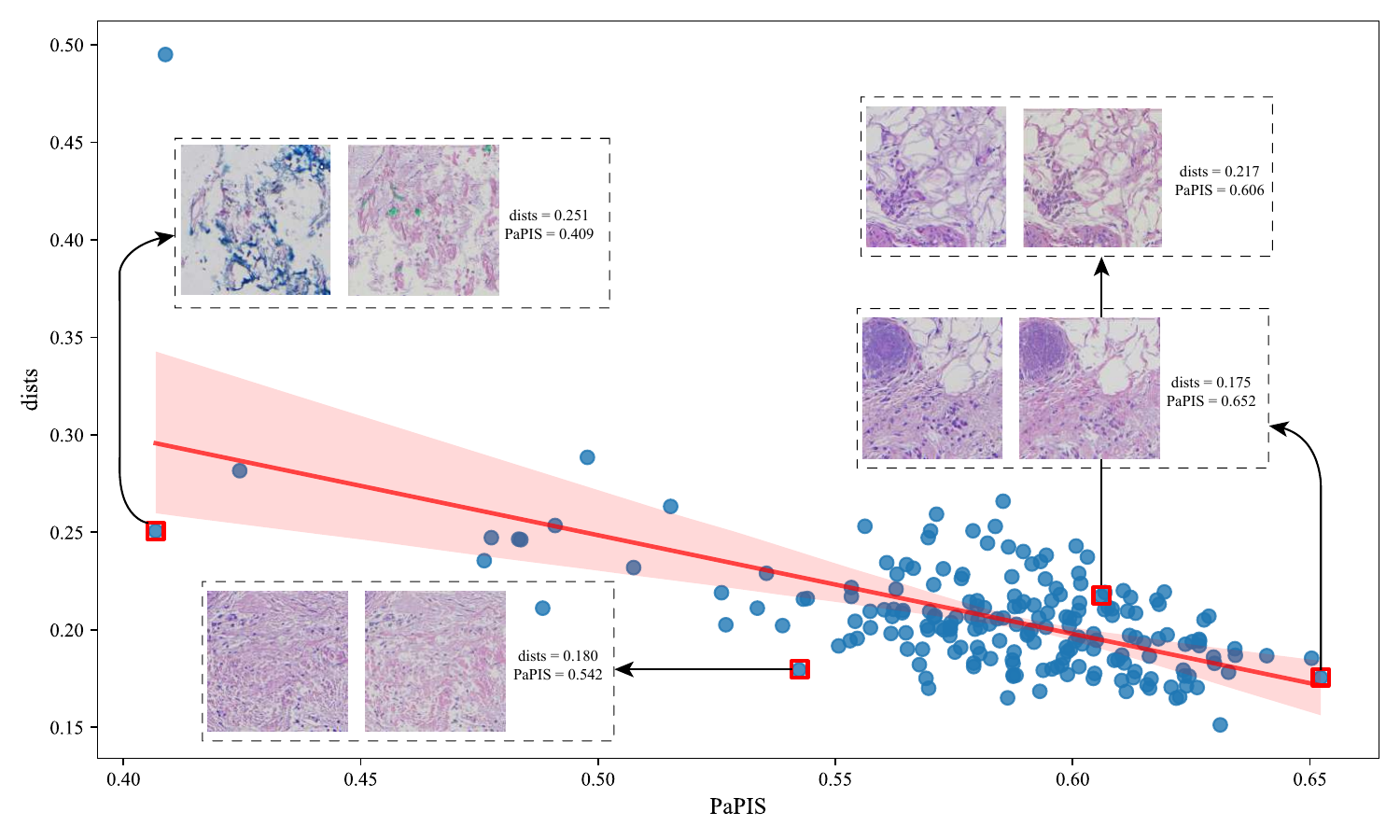}    \caption{Scatter plot comparing DISTS and PaPIS across different image pairs.}
    \label{fig:scatter_plot_dists}
\end{figure}

In this section, PaPIS is compared with other perceptual-based similarity metrics, specifically LPIPS \cite{zhang2018unreasonable} and DISTS \cite{ding2020image}. A case-by-case scatter plot is presented in Figure \ref{fig:scatter_plot_lpips} and \ref{fig:scatter_plot_dists}. At the AH and AL points, both metrics consistently identify images as either high or low in overall quality. However, at the PD points, the images tend to be concentrated in regions with high cellular density, whereas at the TD points, they are primarily located at tissue edges or in areas with sparse cellular distribution. Notably, neither perceptual metric provides explicit information regarding morphological differences in cellular structures. Nevertheless, compared to handcrafted metrics, traditional perceptual metrics exhibit higher consistency with PaPIS, suggesting an improved alignment in assessing histologically relevant features.

\subsection{PaPIS-guided Virtual Staining}

\begin{figure}[htbp]
    \centering
    \includegraphics[width=\columnwidth]{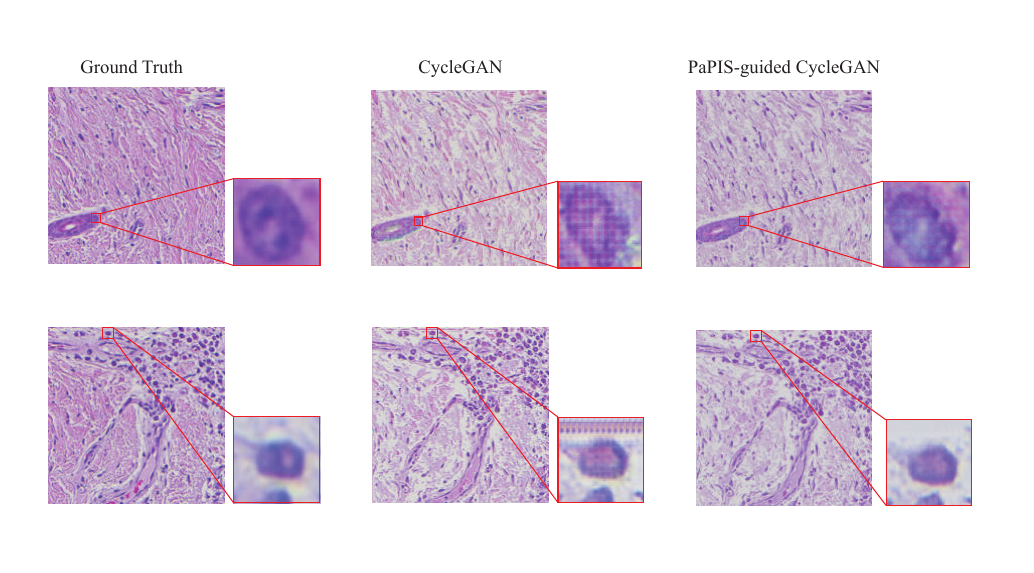}
    \hspace{5cm}
    \caption{Comparison of virtual staining results, showing ground truth images, CycleGAN-generated images, and PaPIS-guided CycleGAN-generated images. The figure illustrates differences in staining quality and structural preservation across the methods. }
    \label{fig:Experiment-PaPIS-guided}
\end{figure}

For the PaPIS-guided virtual staining process, the model is trained following the parameters outlined in Section~\ref{PaPIS-guided}, ensuring a fair comparison of performance. The weight coefficient $\lambda_2$ is set to $2.0$, introducing the $\mathcal{L}_{PaPIS}$ loss term to balance the overall loss function. As shown in Figure~\ref{fig:Experiment-PaPIS-guided}, the PaPIS-guided model enhances the consistency of cellular morphology, including cell position, shape, and size. Additionally, the model effectively mitigates certain image detail losses, leading to a more faithful reconstruction of fine structures. In terms of overall perceptual quality, the generated images exhibit improved resemblance to the ground truth. The PaPIS-guided virtual staining model demonstrates enhanced performance in both histological similarity and high-frequency texture preservation, optimizing pathological feature retention in the generated images.

\subsubsection{Regional Sensitivity Analysis via Heatmap Visualization}

\begin{figure}[htbp]
    \centering
    \includegraphics[width=0.9\columnwidth]{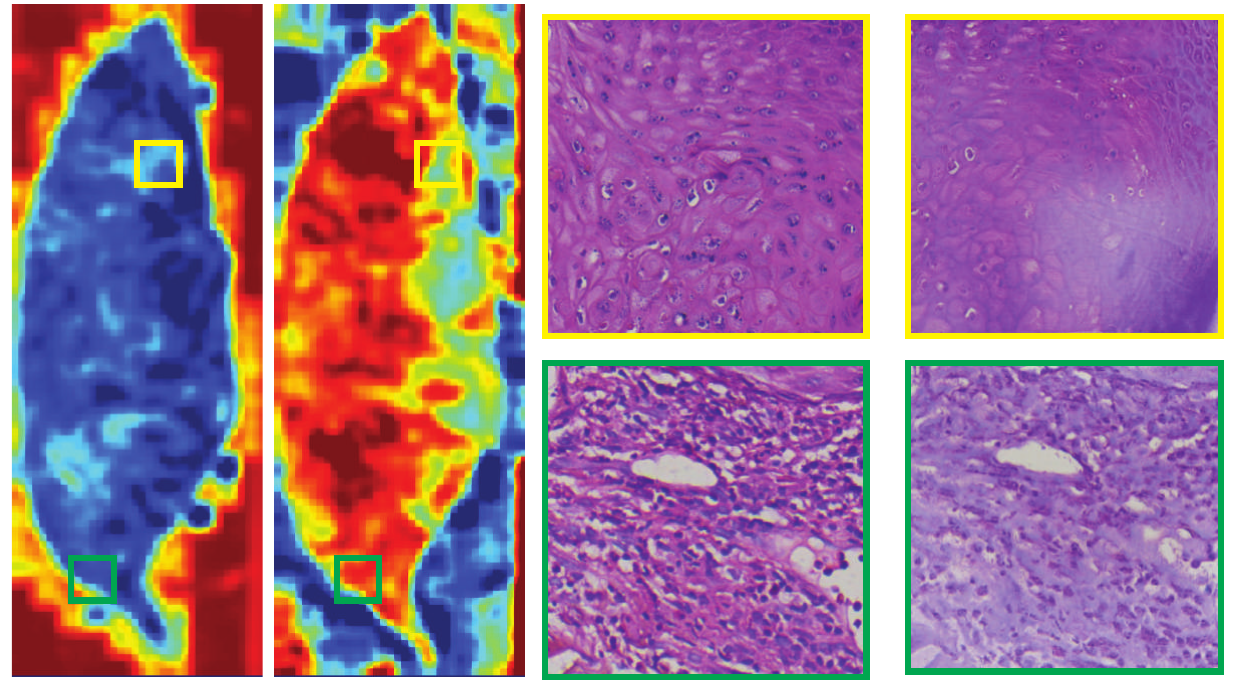}
    \caption{Patch-wise similarity heatmaps of a representative WSI.
    \textbf{Left}: SSIM heatmap; \textbf{Middle}: PaPIS heatmap;
    \textbf{Right}: Magnified views of selected regions. Warmer colors indicate higher similarity.
    Compared to SSIM, which often assigns high similarity scores in low-texture or background areas, PaPIS demonstrates greater sensitivity to morphological and contextual cues.}
    \label{fig:heatmap_papis_ssim}
\end{figure}

To further assess the spatial sensitivity of PaPIS compared to traditional metrics, we conducted a region-wise similarity analysis on a representative whole slide image. The image was divided into non-overlapping \(1024 \times 1024\) patches, and similarity scores were computed using both SSIM and PaPIS. These values were then visualized as spatial heatmaps, as shown in Fig.~\ref{fig:heatmap_papis_ssim}. Warmer colors indicate higher similarity.

\paragraph*{Observations.} From this analysis, we derive the following observations:

\begin{enumerate}
    \item \textbf{Texture Sensitivity vs. Morphological Awareness:} SSIM shows a sharp contrast between background and tissue regions, primarily responding to local texture presence. In contrast, PaPIS produces smoother gradients at tissue boundaries and highlights internal structure variations, suggesting better alignment with morphological integrity.
    
    \item \textbf{Detection of Staining Artifacts:} In certain regions with high SSIM but visually suboptimal staining, PaPIS assigns noticeably lower scores, demonstrating its capacity to penalize virtual staining artifacts even when conventional metrics fail to do so.

    \item \textbf{Intra-Tissue Structural Differentiation:} PaPIS reveals greater variance within tissue interiors than SSIM, which tends to yield flat responses. This indicates that PaPIS is more sensitive to histologically relevant features such as nuclear organization or glandular structure.

    \item \textbf{Sensitivity to Functional Regions:} In regions containing critical micro-anatomical structures (e.g., ducts, tumor nests), SSIM maintains high scores despite morphological degradation. In contrast, PaPIS shows significant score drops, reflecting its better alignment with pathology-relevant structures.

    \item \textbf{Complementary Applications:} The divergence between SSIM and PaPIS highlights potential complementary use cases: SSIM may remain useful for detecting broad texture-based distortions, while PaPIS provides superior guidance for histology-aware evaluation and diagnostic quality control.
\end{enumerate}

These insights underscore the importance of incorporating domain-specific perceptual criteria into similarity metrics. PaPIS enables fine-grained, region-aware evaluation that aligns more closely with expert interpretation and diagnostic relevance, offering a distinct advantage over traditional full-reference IQA methods in virtual staining workflows.

\section{Discussions and Conclusions}

In this study, we introduced PaPIS, a pathology-guided FR-IQA metric tailored for virtual staining evaluation. By leveraging a cell morphology-aware feature extractor and Retinex-based feature decomposition, PaPIS captures both high-frequency nuclear structures and low-frequency illumination characteristics—features highly relevant to histological interpretation. Our experimental results demonstrate that PaPIS achieves superior performance in identifying pathology-relevant differences across image pairs, outperforming traditional handcrafted metrics (e.g., SSIM, PSNR) and perceptual feature-based metrics (e.g., LPIPS, DISTS). Furthermore, the integration of PaPIS as a perceptual loss into a CycleGAN-based virtual staining model improved histological fidelity in generated outputs, suggesting that PaPIS can be used not only as an evaluation tool but also as a training objective.

A notable limitation of the current study is the absence of subjective validation by expert pathologists. While PaPIS shows promising quantitative alignment with structural and morphological features important to histopathology, its clinical validity remains to be established through direct correlation with expert perception and diagnostic utility. This limitation stems from the logistical complexity and resource demands of conducting large-scale reader studies. Nevertheless, we acknowledge this as a critical direction for future work. We plan to design a reader study in collaboration with board-certified pathologists, involving pairwise comparisons of virtual staining outputs across different models and correlation analysis between PaPIS scores and expert ratings. This will help quantify the interpretability and clinical relevance of the proposed metric.

Beyond the current experiments based on PARS imaging, we also anticipate the applicability of PaPIS to other label-free virtual staining modalities such as autofluorescence microscopy, quantitative phase imaging, and hyperspectral imaging. Since PaPIS operates in a feature space that captures cell-level morphology rather than being tied to specific pixel distributions, it is inherently modality-agnostic and can generalize across structurally diverse imaging sources, as long as the target virtual stain preserves histological structures.

Moreover, the modular structure of PaPIS enables generalization across tissue types and staining protocols. While the current implementation uses a feature encoder trained on H\&E-based nuclei segmentation, the framework can be extended by incorporating encoders trained on different tissue domains or fine-tuned using transfer learning. This flexibility opens opportunities for PaPIS to serve as a generalized perceptual metric across various virtual staining scenarios, including different stain types (e.g., Masson's trichrome, IHC) and across species or clinical contexts.

Despite the limitations in subjective validation, the strong quantitative performance and architectural flexibility of PaPIS position it as a valuable tool for automated quality assessment in computational pathology. It offers an interpretable, pathology-aware alternative to conventional IQA metrics and has the potential to guide model selection, training optimization, and quality control in large-scale virtual staining workflows. In future work, we aim to further explore PaPIS's clinical relevance, extend its applicability to broader imaging contexts, and integrate it with human-in-the-loop systems for semi-automated quality assurance.
\backmatter





\bmhead{Acknowledgements}

The authors used OpenAI’s ChatGPT (version GPT-4, accessed May 2025) to assist in improving the English language and clarity of the manuscript during the revision stage. All intellectual content was written and verified by the authors.

\bibliography{sn-bibliography}

\end{document}